\documentclass[twocolumn,aps,prd,superscriptaddress,showpacs,floatfix]{revtex4}

\usepackage{epsfig,bm,feynmf}
\usepackage{amsmath}
\usepackage{graphicx}

\usepackage[normalem]{ulem} 
\renewcommand\sout{\bgroup \color{red} \ULdepth=-.5ex \ULset}
\usepackage[dvips]{color} 

\def\ltap{\ \raise.3ex\hbox{$<$\kern-.75em\lower1ex\hbox{$\sim$}}\ }
\def\gtap{\ \raise.3ex\hbox{$>$\kern-.75em\lower1ex\hbox{$\sim$}}\ }

\begin{document}



\title{Properties of $J/\psi$ at $T_c$: QCD second-order Stark effect}

\author{Su Houng Lee}
\email{suhoung@phya.yonsei.ac.kr}
\author{Kenji Morita}
\email{morita@phya.yonsei.ac.kr}
\affiliation{Institute of Physics
and Applied Physics, Yonsei University, Seoul 120-749, Republic of
Korea}


\begin{abstract}
 Starting from the temperature dependencies of the energy density and
 pressure from lattice QCD calculation, we extract the temperature
 dependencies of the electric and magnetic condensate near $T_c$.
 While the magnetic condensate hardly changes across $T_c$, we
 find that the electric condensate increases abruptly above
 $T_c$.  This induces a small but an equally abrupt decrease in
 the mass of $J/\psi$, which can be calculated through
 the second-order Stark effect.  Combining the present result with the
 previously determined QCD sum rule constraint, we extract the thermal
 width of $J/\psi$ above $T_c$, which also increases fast.  These
 changes can be identified as the critical behavior of $J/\psi$ across
 $T_c$ associated with the phase transition.  We find that the mass
 shift and width broadening of $J/\psi$ at 1.05 $T_c$ will be
 around $-100$ MeV and 100 MeV respectively.
\end{abstract}



\maketitle
Since the pioneering work by Hashimoto et al.~\cite{Hashimoto86} and the
seminal work by Matsui and Satz~\cite{Matsui86}, many experimental and
theoretical works have been performed in the physics of $J/\psi$
suppression in heavy ion collisions.   The subject has recently evolved
into a new stage as lattice calculations using maximum entropy methods
(MEM) found the peak structure to survive up to almost
2$T_c$~\cite{Hatsuda04,Datta04}, which was speculated
before~\cite{Hansson87} as lattice calculations showed that
non-perturbative nature of QCD persists well above
$T_c$~\cite{Polonyi87,Lee89}.  This suggests that the sudden
disappearance of $J/\psi$ is not the direct signature of QGP formation.
Indeed, recent results from RHIC on the suppression factors at different
rapidity and higher $p_T$ seem ever more confusing~\cite{Phenix}.
Hydrodynamic calculations suggest that the initial temperature of the
QGP formed at RHIC is in the order of $2T_c$ and will last for 3 to 4
fm/$c$~\cite{Heinz,Morita-hydro}.  Therefore, considering the  formation
time of charmonium states after the creation of $\bar{c}c$ pair, it is
crucial to know the detailed properties of $J/\psi$ near $T_c$
to fully understand the suppression and/or enhancement of $J/\psi$ in
heavy ion collisions.  Unfortunately, the present resolutions of the
peak structure of $J/\psi$ from the lattice calculations based on MEM
are far from satisfactory~\cite{Jakovac06}.  In fact, the peak is too
broad to even discriminate between $J/\psi$ from $\psi'$.    Moreover,
the temperature region between $T_c$ to 2$T_c$ is
known to be strongly interacting and therefore a non-perturbative method
has to be implemented to consistently treat the charmonium at this
temperature region. In a previous work~\cite{Morita_prl}, we have
implemented QCD sum rules to
investigate the properties of $J/\psi$ near $T_c$.   Although the
results were non-perturbative, only a constraint on the combined mass
decrease and width increase could be obtained.  Here, we  point out that
the critical behavior of QCD phase transition, could be identified with
a critical behavior of electric condensate at $T_c$, and then
by making use of the QCD second order Stark effect, show that such
critical behavior can be translated to a sudden change in the mass of
$J/\psi$ across $T_c$.

\begin{figure}[tb]
\begin{center}
\includegraphics[width=3.375in]{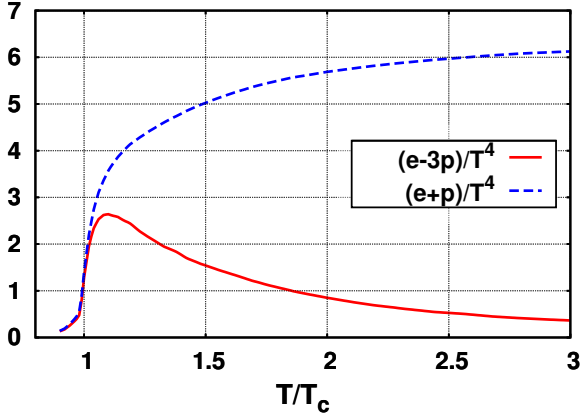}
\end{center}
\caption{$M_0=(\varepsilon-3p)$ and $M_2=(\varepsilon+p)$  divided by $T^4$
 as functions of temperature.
}
\label{g0g2}
\end{figure}

We begin by characterizing the properties of the
strongly interacting quark-gluon plasma (sQGP) of the pure gluon theory
across $T_c$ in terms of local operators.  This is accomplished by
making use of the energy-momentum tensor, which has a symmetric
traceless part and a trace part via the trace anomaly,

\begin{eqnarray}
T_{\alpha \beta}= -{\cal S T} ( G^a_{ \alpha \mu}G_\beta^{a   \mu})+ \frac{g_{\alpha \beta}}{4} \frac{\beta(g)}{2 g} G^a_{\mu \nu }G^{a \mu \nu }.
\label{emt}
\end{eqnarray}
Here, $a$ and $\alpha, \beta$ are the color and  Lorentz indices respectively.
The temperature dependence of the two independent parts of the
energy-momentum tensor can be obtained from the lattice measurement of
energy density and pressure at finite temperature.
\begin{equation}
\langle T_{\alpha \beta} \rangle_T = (\varepsilon+p)\left(u_\alpha u_\beta -				    \frac{1}{4}g_{\alpha \beta}\right)
 + (\varepsilon-3p)\frac{g_{\alpha\beta}}{4}.\label{eq:e-m-t}
\end{equation}
Here, $u_\alpha$ is the four velocity of the heat bath.
Therefore the temperature dependencies of gluonic operators can be identified with the pressure and energy density.  To leading order in coupling, we can identify the trace part
$-\frac{11}{8}  [\langle \frac{\alpha_{\rm s}}{\pi}G^{a}_{  \mu \nu}
G^{a \mu \nu} \rangle_T - \langle \frac{\alpha_{\rm s}}{\pi}G^{a}_{  \mu \nu}
G^{a \mu \nu} \rangle_0 ] = M_0(T)$ and the non-trace part $\langle - {\cal S T} G^a_{\alpha \mu}G_\beta^{a \mu } )\rangle_T =
(u_\alpha u_\beta - \frac{1}{4}g_{\alpha \beta} )  M_2(T)$,
where~\cite{Morita_prl},
\begin{eqnarray}
M_0(T) & = &  (\varepsilon-3p), \nonumber \\
M_2(T) & = & (\varepsilon+p), \label{gep}
\end{eqnarray}
and  $ \left\langle \frac{\alpha_{\rm s}}{\pi}G^{a}_{  \mu \nu}
G^{a \mu \nu} \right\rangle_0 \equiv G_0^{\rm vac}$ is the scalar gluon
condensate in the vacuum.  The lattice gauge theory result for
$\varepsilon$ and $p$ in
the pure SU(3) gauge theory was obtained from Ref.~\cite{Boyd96}.
Figure \ref{g0g2} shows changes of $M_0$ and $M_2$ from their vacuum value scaled by their asymptotic temperature dependence
of $T^4$.  One notes that while $M_2$, which is also proportional to
entropy density times temperature, moderately reaches the asymptotic
temperature dependence, $M_0$, also known as the interaction measure or
the gluon condensate, suddenly increases and then decreases at higher
temperature.  The strongly interacting nature of QGP is related to the
large interaction measure~\cite{Kharzeev-Karsch08}, which takes its
maximum value at around 1.1$T_c$.  It should be noted that the
temperature dependence of the gluon condensate $M_0$ extracted here
includes both the perturbative and the non-perturbative and thus the
full temperature dependence.  The sudden change near $T_c$ is
dominated by the sudden decrease of the non-perturbative part, which
reduces to about half of its vacuum value~\cite{Lee89}, while at higher
temperatures it is dominated by the perturbative
contributions~\cite{Boyd96,Miller06}.

For the heat bath at rest, one can rewrite the thermal expectation
values of the dimension four operators of the energy-momentum tensor in
Eq.~\eqref{emt} in terms of electric and magnetic
condensate~\cite{Luke-Manohar}.  This is possible after making the following identification.
\begin{eqnarray}
\left\langle
 \frac{\alpha_{\rm s}}{\pi}{\cal S T} (G^a_{\alpha \mu}G_\beta^{a \mu })
 \right\rangle_T
\equiv
\frac{\alpha_{\rm s}(T)}{\pi}
\left\langle {\cal S T} (G^a_{\alpha \mu}G_\beta^{a \mu })\right\rangle_T.
\end{eqnarray}
The scale dependence of the matrix element is transferred to the
coupling constant.  Therefore, we additionally need to know the
temperature dependence of the coupling constant $\alpha_{\rm s}(T)$.
Since we will be using the matrix element in the operator product
expansion (OPE) with the separation scale relevant for the heavy bound
state,   we
will use the temperature dependent running coupling constant extracted
from the lattice computation of the heavy quark free energy
\cite{alpha-T}.  Then we find,
\begin{align}
\left\langle \frac{\alpha_{\text{s}}}{\pi} \boldsymbol{E}^2
\right\rangle_T-\left\langle
\frac{\alpha_{\text{s}}}{\pi}\boldsymbol{E}^2 \right\rangle_0 & =  \frac{2}{11}M_0(T) +\frac{3}{4}\frac{\alpha_{\rm s}(T)}{\pi} M_2(T),
 \label{eq:defe2}\\
\left\langle \frac{\alpha_{\text{s}}}{\pi} \boldsymbol{B}^2
\right\rangle_T
-\left\langle \frac{\alpha_{\text{s}}}{\pi}\boldsymbol{B}^2 \right\rangle
  &=  - \frac{2}{11}M_0(T) + \frac{3}{4} \frac{\alpha_{\rm s}(T)}{\pi} M_2(T). \label{eq:defb2}
\end{align}

\begin{figure}[htb]
\begin{center}
\includegraphics[width=3.375in]{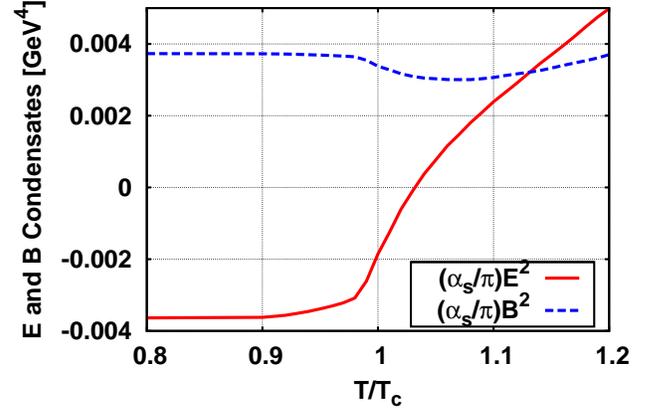}
\end{center}
\caption{Electric and magnetic condensate near $T_c$ as functions of
temperature.}
\label{e2b2}
\end{figure}

Figure \ref{e2b2} shows the temperature dependence of
$\langle\frac{\alpha_{\rm s}}{{\pi}} \boldsymbol{E}^2 \rangle_T$ and
$\langle\frac{\alpha_{\rm s}}{{\pi}} \boldsymbol{B}^2 \rangle_T$.
One notes that there is a sudden increase in the electric condensate
$\langle \frac{\alpha_{\rm s}}{{\pi}}\boldsymbol{E}^2 \rangle_T$, while
the magnetic condensate
$\langle \frac{\alpha_{\rm s}}{{\pi}}\boldsymbol{B}^2 \rangle_T$ hardly
changes above $T_c$.   This can be related to the fact that the
area law behavior of the space-time Wilson loop changes to the perimeter
law above $T_c$, while that of the space-space Wilson loop
retains the area law behavior even above
$T_c$~\cite{Polonyi87}. The connection comes in as the non
perturbative behavior
of a rectangular Wilson loop in the $S_1 S_2$ direction can be related
to the non-vanishing gluon condensate
$\langle \frac{\alpha_{\rm s}}{{\pi}} G_{S_1 S_2}^2 \rangle $ via the
operator product expansion~\cite{Shifman80}.  Hence, one can conclude
that critical behaviors of QCD phase transition can be related to the
sudden change in the electric condensate.   Such local changes will
induce critical  behavior of a heavy quark system such as the $J/\psi$
across the phase transition, which can be obtained through the QCD
second-order Stark effect.


The perturbative QCD formalism for calculating the interaction between
heavy quarkonium and partons was first developed by
Peskin~\cite{Peskin79,BP79} in the non-relativistic limit.   The formula
for the mass shift reduces to the second-order Stark effect in QCD,
which was used previously to calculate the mass shift of charmonium in nuclear
matter~\cite{Luke-Manohar}. The information needed from the medium is
the electric field square.  As the dominant change across the
phase transition is the electric condensate, one notes that the
second-order Stark effect is the most natural formula to be used across
the phase transition.

The second-order Stark effect for the ground state charmonium with
momentum space wave function normalized as
$\int \frac{d^3p}{(2 \pi)^3} |\psi(\boldsymbol{p})|^2=1$ is as follows,
\begin{eqnarray}
\Delta m_{J/\psi} & = &
 - \frac{1}{18} \int_0^\infty dk^2
 \left| \frac{\partial \psi(k)}{\partial k}
 \right|^2
 {k \over k^2/m_c+ \epsilon}
 \left\langle \frac{\alpha_{\rm s}}{\pi} \Delta\boldsymbol{E}^2
	\right\rangle_T \nonumber \\
& = & -\frac{ 7\pi^2}{18} \frac{a^2}{\epsilon}
 \left\langle \frac{\alpha_{\rm s}}{\pi} \Delta\boldsymbol{E}^2 \right\rangle_T , \label{stark}
\end{eqnarray}
where $k=|\boldsymbol{k}|$ and $\langle
\frac{\alpha_{\text{s}}}{\pi}\Delta \boldsymbol{E}^2 \rangle_T$
denotes the value of change of the electric condensate from its vacuum
value.
The second line is obtained for the Coulomb wave function.
Here, $\epsilon$ is the binding energy and $m_c$ the charm quark mass.
These parameters are fit to the size of the wave function obtained in
the Cornell potential model~\cite{Eichten_PRD21}, and to the mass of $J/\psi$
assuming it to be a Coulombic bound state in the heavy quark
limit~\cite{Peskin79}.  The fit
gives $m_c=1704$ MeV, $a=0.271$ fm and $\alpha_s=0.57$.   Few comments
are in order.  The minus sign in Eq.~\eqref{stark} is a model
independent result and follows from the fact that the second-order Stark
effect is negative for the ground state.  The factor of Bohr radius
square $a^2$ follows from the dipole nature of the
interaction, and the binding energy $\epsilon$ from the inverse propagator,
characterizing the separation scale~\cite{Peskin79,OKL02}.   Therefore,
the actual value of the mass shift does not depend much on the form of
the wave function as long as the size of the wave function is fixed.
The Bohr radius used in our calculation corresponds to
$\langle r^2 \rangle^{1/2}=0.47$ fm, which is the  size of a more
realistic wave function in the Cornell potential~\cite{Eichten_PRD21}.
Therefore, the correction coming from using a more realistic wave
function should be small.

\begin{figure}[tb]
\begin{center}
\includegraphics[width=3.375in]{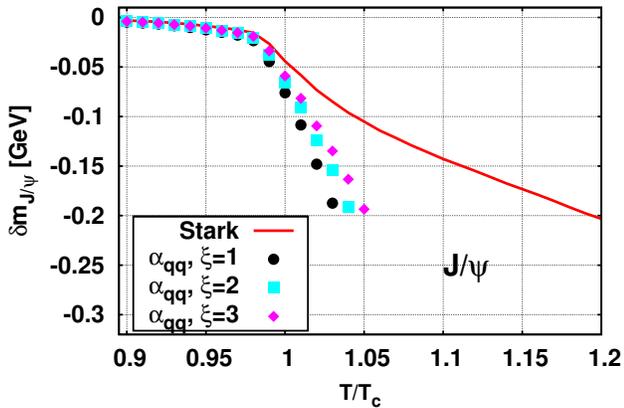}
\end{center}
\caption{Mass shift from the second-order Stark effect (solid line) and
 the maximal mass shift obtained from QCD sum rules from
 Ref.~\cite{Morita_prl} (points).
}
\label{mass-shift}
\end{figure}

The solid line in Fig.~\ref{mass-shift} shows the mass shift obtained
from the second-order Stark effect.  The second-order Stark effect
formula is based on the operator product expansion (OPE) for bound
state. As mentioned before, the formalism was first established by
Peskin in 1979 and the separation scale is the binding
$\epsilon=mg^4$. A more systematic derivation was developed recently by
Brambila et al. ~\cite{Brambila05} for the bound state.  Here, the
relevant scales are $mv$ and $mv^2$, where the former is related to the
potential $1/r$ and later the kinetic energy $p^2/m$.   This scale
$mv^2$ is the separation scale so
that for effects with typical momentum larger than the separation scale
should be taken into account through resummed perturbation by solving
the Schr\"{o}dinger equation\cite{OKL02},
while that with smaller scale should be taken into account through the
operator product expansion.   The operators contain the non-perturbative
physics of QCD, which is typically of order $\Lambda_{\text{QCD}}$.  Therefore
the OPE for the bound state works best when $mv^2 \gg \Lambda_\text{QCD}$~\cite{Brambila05}.  While there are concerns that this
condition is marginal for charmonium, the approach should provide a
quantitative description.   Now the question is, which approach should
one take for the thermal interactions near $T_c$. Obviously, the effects
of finite temperature involves new scales like the color
screening. However, as has been known for some time, the
temperature region from $T_c$ to 2 $T_c$, is known to be strongly
interaction and can not be described by resumed perturbation~\cite{Blaizot99}.
Therefore, the approach we want to take is to calculate the non
perturbative temperature effect to the mass shift through the
operator product expansion. The leading order contribution in
this approach is coming from $\left\langle \frac{\alpha_{\rm s}}{\pi}
\Delta\boldsymbol{E}^2 \right\rangle_T$ as is given in
Eq.~\eqref{stark}, whose
temperature dependence we extract directly from the lattice.
The arguments for convergence of higher dimensional operators
in our approach are twofold. First, we will restrict to the
temperature region where the change in $\left\langle \frac{\alpha_{\rm
s}}{\pi} \Delta\boldsymbol{E}^2 \right\rangle_T$ is smaller than the
vacuum value of $\left\langle \frac{\alpha_{\rm s}}{\pi}
\boldsymbol{E}^2 \right\rangle_0$ itself. We believe that then the OPE
is under control as has been verified by the typical QCD sum rule
approaches for heavy quark system in the vacuum. As can be seen
in Fig.~\ref{e2b2}, this condition restricts our
applicability to 1.05$T_c$, as in the QCD sum rule approach at finite
temperature~\cite{Morita_prl,Morita_prc}. Second, a more direct evidence
comes from the next term in the OPE correction, which comes from
magnetic condensate. However, as can be seen from Fig.~\ref{e2b2}, the
changes of $\left\langle \frac{\alpha_{\rm s}}{\pi}
\Delta\boldsymbol{B}^2 \right\rangle_T$ and hence the next term in the
OPE should be small up to 1.05$T_c$ and slightly beyond.
Therefore, we can conclude that the second-order Stark effect should be
valid near $T_c$.   As can be seen in Fig.~\ref{mass-shift},
the results from second-order Stark effect shows that the mass reduces
abruptly above $T_c$ and becomes smaller by about 100 MeV at
1.05$T_c$, reflecting the critical behavior of the QCD phase
transition.

\begin{figure}[tb]
\begin{center}
\includegraphics[width=3.375in]{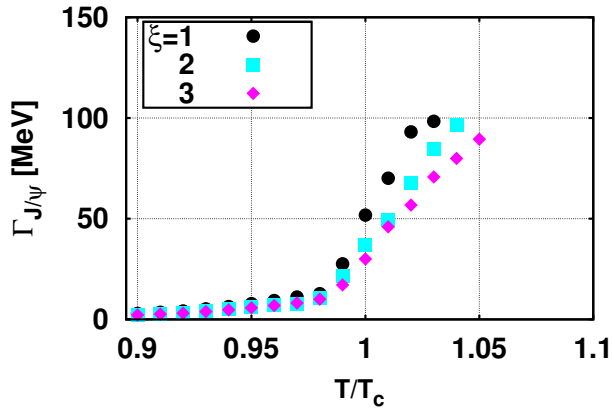}
\end{center}
\caption{Thermal width of $J/\psi$ obtained from the second-order Stark
 effect and QCD sum rule constraint.}
\label{thermal-width}
\end{figure}

We put the present result in perspective with a non-perturbative result obtained
before using the QCD sum rules~\cite{Morita_prl,Morita_prc}.  The points
in Fig.~\ref{mass-shift} represent the maximum mass shift obtained in
Refs.~\cite{Morita_prl,Morita_prc}.  As can be seen in the figure, the mass
shift obtained from the second-order Stark effect is almost the same as
the maximum mass shift obtained in the sum rule up to $T_c$ and then
becomes smaller. The mass shift at $T_c$ is about $-50$ MeV.
In the QCD
sum rules, only a constraint for the combined mass shift and thermal
width of $-\Delta m+\Gamma_T\simeq 80+17(T-T_c)$ MeV could be obtained
within the temperature range from $T_c$  to $1.05T_c$.
Therefore, the difference between the Stark effect and the maximum mass
shift obtained from QCD sum rules above $T_c$ in
Fig.~\ref{mass-shift} could be
attributed to the non-perturbative thermal width at finite temperature.
In Fig.~\ref{thermal-width}, we plot the thermal width obtained from
combining the QCD sum rule constraint with the mass shift
obtained from the QCD second-order Stark effect.\footnote{We have
improved the QCD sum rule calculation by taking into account running
mass effect properly. This changes the result of $\xi \neq 1$ from
Ref.~\cite{Morita_prc}.}   As can be seen in the
figure, the thermal width at 1.05$T_c$ becomes as large as 100 MeV.
Such width slightly above $T_c$ is larger than that estimated from a
perturbative LO and NLO QCD method~\cite{Park07,Song07}, but smaller
than a recent phenomenological estimate~\cite{Mocsy}.   The mass of
quarkonium at finite temperature was also  investigated in the potential
models~\cite{Alberico06}, where the mass was found to decrease at high
temperature.  However, the detailed potential has to be extracted from
the lattice at each temperature and hence identifying the critical
behavior near $T_c$ will be difficult.

Finally, we comment that while the present result is obtained using lattice
calculation in the pure gauge theory, including dynamical quark will not
greatly modify the result.
This follows from noting that the
lattice result for the temperature dependence of pressure
$p/T^4$ is independent of the number of flavors if scaled to their
corresponding ideal gas limit~\cite{Karsch01}.  Moreover, as was shown
in Ref.~\cite{lee_qm08}, $G_0(T)$ extracted from a recent full
lattice calculation of the interaction measure~\cite{cheng08} after
subtracting the quark contributions, and then dividing by a factor of
$(1+\frac{5}{12}n_f)$ appearing in the beta function, shows that the
change of the magnitude near $T_c$ is remarkably similar to
that of the pure gauge theory.  Hence the main input for our result does
not change much even in the presence of dynamical quarks.

In summary, we have shown that the sudden increase in the energy density
across the phase transition, which is a characteristic behavior of the QCD
phase transition independent of the flavor, can be translated to a rapid
increase in the electric condensate slightly above $T_c$.
Using the QCD second-order Stark effect, this translates into an equally
sudden decrease in the mass of $J/\psi$, which is around $-50$ MeV and
$-100$ MeV respectively at $T_c$ and 1.05$T_c$.
Combining with a QCD sum rule
constraint, we obtain the thermal width of $J/\psi$ slightly above
$T_c$, and found it to be larger than previous perturbative estimates,
and becoming as much as 100 MeV at 1.05$T_c$.  Hence, one can conclude
that the critical behavior of $J/\psi$ at $T_c$ is not its sudden
disappearance, but rather the abrupt changes of its mass and width. The
mass shift is probably too small to be detected with present resolutions
at RHIC.  However, with the expected upgrades at RHIC and plans at LHC,
such direct measurement could be possible. Indeed, the mass resolution
of $J/\psi$ for dimuon channel at LHC is 35 MeV for the CMS
detector~\cite{CMS_res} and around 70 MeV for ALICE~\cite{ALICE_res} and
ATLAS~\cite{ATLAS_res}. It might be better
for dielectron channel. Furthermore, the mass shift could also influence
production rates within the statistical model~\cite{PBM}. The large
width of 100--150 MeV already at 1.05$T_c$ suggest that
while the maximal entropy method shows a $J/\psi$ peak structure
surviving  up to 2$T_c$, the actual formation
at heavy ion collisions might only be possible at lower temperatures where the width of the $J/\psi$ becomes equal to its binding.



This work was supported by  the Korean Ministry of
Education through the BK21 Program and
KRF-2006-C00011.

\end{document}